# Interpretable domain knowledge enhanced machine learning framework on axial capacity prediction of circular CFST columns

Dian Wang[a], Zhigang Ren[a]*, Gen Kondo[b]*

a School of Civil Engineering and Architecture, Wuhan University of Technology, No. 122 Luoshi Road, Wuhan 430070, China
b Department of Civil and Environmental Engineering, University of California, Berkeley, CA 94720, USA

* Corresponding authors.
E-mail address: renzg@whut.edu.cn (Z. Ren), kondogen@berkeley.edu (G. Kondo)

## Abstract:

This study introduces a novel machine learning framework, integrating domain knowledge, to accurately predict the bearing capacity of CFSTs, bridging the gap between traditional engineering and machine learning techniques. Utilizing a comprehensive database of 2621 experimental data points on CFSTs, we developed a Domain Knowledge Enhanced Neural Network (DKNN) model. This model incorporates advanced feature engineering techniques, including Pearson correlation, XGBoost, and Random tree algorithms. The DKNN model demonstrated a marked improvement in prediction accuracy, with a Mean Absolute Percentage Error (MAPE) reduction of over 50% compared to existing models. Its robustness was confirmed through extensive performance assessments, maintaining high accuracy even in noisy environments. Furthermore, sensitivity and SHAP analysis were conducted to assess the contribution of each effective parameter to axial load capacity and propose design recommendations for the diameter of cross-section, material strength range and material combination. This research advances CFST predictive modelling, showcasing the potential of integrating machine learning with domain expertise in structural engineering. The DKNN model sets a new benchmark for accuracy and reliability in the field.

Keywords: Concrete-filled steel tube, Machine learning, Axial compression capacity, Domain knowledge, Neural Network, SHAP

# 1. Introduction

Concrete-filled steel tube (CFST) structures are extensively utilized in engineering for their high strength, stability, and durability. However, predicting the bearing capacity of CFST members under load accurately is challenging, influenced by factors like material properties, structural forms, and environmental conditions. Conventional methods, based on empirical formulas and experimental data, struggle with complex non-linear material behavior and intricate structural configurations, limiting their accuracy and applicability [1]. Advanced analytical techniques, such as neural networks (NNs), have gained attention as a preferred method for the prediction of bearing capacity. NNs' ability to learn complex nonlinear relationships from large datasets makes them a compelling choice for accurately calculating CFSTs' bearing capacity, drawing significant research interest [2–4].

While NNs exhibit remarkable adaptability and predictive capabilities, they have certain limitations. Firstly, NN models are considered black-box models, making it difficult to interpret their decision-making process. This lack of transparency is problematic in engineering, where practical applications demand explanations and validation of predictions [5]. Secondly, traditional data-driven machine learning (ML) methods may yield unreliable outputs in data spaces lacking information, affecting generalization ability due to data distribution and sample size limitations [6]. A summary of recent papers related to NN models in the field of CFST structures is presented in Table 1. Addressing these issues to advance ML in CFST structures requires incorporating domain knowledge and practical application requirements before implementing ML, ensuring meaningful interpretation and utilization of prediction results. In recent years, domain knowledge-based optimization and prediction methods for NNs, such as Physics-Informed NNs (PINN) and Physics Model-based Neural Networks (Phy-Net), have become prominent in the scientific research [7]. These methods leverage domain knowledge as prior knowledge, combined with extensive experimental data for training, leading to accurate problem-solving and model optimization. In the field of engineering, the integration of domain knowledge into NNs is known to as Knowledge-based NNs [5], Domain Knowledge-aided ML [8], or Domain Adapted Neural Networks (DANN) [9].

Table 1 Summary of the literatures

| | Ref | [10] | [11] | [12] | [12] | [13] | [14] | [3] | [15] | [16] | [17] | [18] | [19] | [20] | This work |
|---|---|---|---|---|---|---|---|---|---|---|---|---|---|---|---|
| Network structure | Layer | 1 | 1 | 1 | 1 | 5 | 1 | 1 | 1 | 2 | 1 | 2 | 1 | 1 | 5 |
| | Neurons Number | 10 | 11 | 10 | 10 | 32 | 12 | 10 | 11 | 5 | 5 | 4/3 | 7 | 10 | 32 |
| Activation Function | Hidden layer | tansig | tansig | tansig | tansig | satlins | tansig | logistic | tansig | tansig | tansig | tansig/logsig | logsig | tansig | relu |
| | Output layer | purelin | purelin | purelin | purelin | purelin | purelin | relu | purelin | purelin | purelin | logsig | logsig | purelin | relu |
| Sample size | | 305 | 646 | 149 | 410 | 1,245 | 2,045 | 509 | 768 | 1305 | 768 | 150 | 633 | 268 | 2621 |
| Feature range | $f_y$ (MPa) | 226~466 | 184.78~834.27 | 186~863.00 | 184.8~1153 | 178.28~853 | ≤229.88 | 181.89~681.89 | 235~460 | 179~853 | 235.0~460.0 | 186–853 | 178.28~853 | 200.2~853 | 178.28~1153 |
| | $f_c'$ (MPa) | 17.91~140.30 | 10.00~198 | 18.03~193.3 | 23~188.1 | 7.7938~177.83 | ≤1233 | 9.17~121.60 | 100~200 | 8~185 | 100.0~200.0 | 18~193 | 9.9~108 | 18~106 | 6.41~200 |
| | L/D | ≤6 | | | | 0.81~51.48 | | | | 0.81~51.48 | | 1.8~4.9 | | | 0.33~51.48 |
| Features | D | | | ✓ | ✓ | ✓ | ✓ | ✓ | ✓ | ✓ | ✓ | ✓ | ✓ | ✓ | ✓ |
| | t | | | ✓ | ✓ | ✓ | ✓ | ✓ | ✓ | ✓ | ✓ | ✓ | ✓ | ✓ | |
| | $f_y$ | ✓ | ✓ | ✓ | ✓ | ✓ | ✓ | ✓ | ✓ | ✓ | ✓ | ✓ | ✓ | ✓ | |
| | $f_c'$ | ✓ | ✓ | ✓ | ✓ | ✓ | ✓ | ✓ | ✓ | ✓ | ✓ | ✓ | ✓ | ✓ | ✓ |
| | $f_u$ | | | | | | | | | ✓ | ✓ | | | | |
| | L | ✓ | | ✓ | ✓ | | ✓ | ✓ | ✓ | ✓ | ✓ | ✓ | ✓ | ✓ | |
| | $A_s$ | ✓ | ✓ | | | | | | | | | | | | ✓ |
| | $A_c$ | ✓ | ✓ | | | | | | | | | | | | ✓ |
| | $E_c$ | | | ✓ | | | | | | | | ✓ | | | |
| | $E_s$ | | | ✓ | | | | | | | | ✓ | | | |
| | D/t | | | ✓ | ✓ | | | | | | | ✓ | | | |
| | λ | | | ✓ | ✓ | | | | | | | | | | |
| | L/D | | | | ✓ | | | | | | | | | | |
| | ξ | | | | ✓ | | | | | | | | | | |
| | t/D | | | | | ✓ | | | | | | ✓ | | | |
| | $α_{sc}$ | | | | | | | | | | | | | | ✓ |
| | $N_{u0}$ | | | | | | | | | | | | | | ✓ |
| | C | | | | | | | | | | | | | | ✓ |
| | $N_s$ | | | | | | | | | | | | | | ✓ |
| | $V_c$ | | | | | | | | | | | | | | ✓ |
| | $V_s$ | | | | | | | | | | | | | | ✓ |

The central idea of this research concept emphasizes the enhancement of NNs through the integration of domain-specific knowledge with advanced ML methodologies. This approach aims to augment interpretability, bolster reliability, refine predictive accuracy, and expand extrapolation capabilities, thereby positioning it as a pivotal direction in contemporary research.

This methodology leverages domain knowledge as a pivotal constraint to steer the learning mechanisms of NNs, ensuring the generation of outputs that are physically significant. In scholarly discourse, domain knowledge is broadly defined, encompassing elements such as relevant features, concepts, taxonomies, empirical rules, logical constraints, probability and mathematical distributions, causal relationships, and others, as delineated in [10]. The application of domain knowledge to NNs presents a methodological advancement, optimizing the utilization of limited data, curbing data demands, and minimizing data processing costs. Additionally, the integration of domain knowledge imparts an enriched foundation of prior information, thereby facilitating the model's enhanced generalization to unfamiliar datasets and novel problems. Such integration not only bolsters the trustworthiness of the models but also deepens their interpretative potential, making them more amenable to thorough examination. Consequently, leveraging this approach in the context of predicting the strength of CFSTs is posited to offer substantial benefits, marking a significant stride in this realm of study.

The integration of domain knowledge into NNs predominantly encompasses constraint-based and guidance-based methodologies. Constraint-based approaches incorporate physical laws within the parameters or loss functions of NNs, effectively narrowing the solution space of the model. Conversely, guidance-based methods employ domain knowledge as a priori information for the input data, steering the learning and predictive processes of the model. This includes emulating the physical behaviors of structures, isolating physical features, and leveraging these as inputs for neural networks to predict structural performance, as documented in references [11,12]. Significant initial investigations have been undertaken in this area. Fernández introduced a novel physics-guided Bayesian neural network, integrating physically-based components into various layers of the neural network to enhance extrapolation capabilities, a notable improvement over existing algorithms [6]. Zhang developed a physics-guided Convolutional Neural Network (PhyCNN) for data-driven seismic response modeling of structures. This innovative approach involved training a deep PhyCNN model based on limited seismic input-output datasets and physical constraints to establish an alternative model for the structural response prediction [13]. Cheng et al. formulated a physically supervised, interpretable ML approach for seismic failure modes of reinforced concrete columns. This first combined insights from the mechanism of column seismic failure modes, experimental data patterns, and engineering experience to reveal the physical laws between key feature parameters of columns and seismic failure modes. Then, they used multi-class logistic regression to propose an interpretable model for seismic failure modes of RC columns [14]. However, it should be noted that ML models based on physical models have not yet been applied to the study of concrete-filled steel tube components. This indicates a potential area for further research and development in the field of CFSTs.

In addition, SHAP (SHapley Additive Explanations) is another method used to enhance the interpretability of ML models by estimating the individual contributions of input features to the model's output. SHAP considers the average contribution of each feature across all possible feature subsets, enabling the estimation of each feature's importance [15]. For the prediction of the bearing capacity of CFSTs, SHAP can be employed to explain the model's contributions of each input feature, such as steel tube diameter, thickness, and concrete strength, among others. This analysis is instrumental in comprehending the decision-making process of the model and gaining a deeper insight into its performance and prediction outcomes. Additionally, based on the analysis of feature importance, feature selection can be performed to reduce model complexity and improve its accuracy [16]. Thus, SHAP offers valuable insights for both model interpretation and optimization in the domain of CFST structural engineering.

In this study, a model based on Domain Knowledge Enhanced Neural Networks (DKNN) is proposed for predicting the bearing capacity of CFSTs. The research aims not only to enhance current data-driven ML models but, more importantly, to embed domain knowledge into artificial intelligence technique to gain a deeper understanding of AI models and their potential engineering benefits. To achieve this goal, an ML framework that incorporates civil engineering expertise at each crucial stage, from data analysis, preprocessing and feature selection to model training and evaluation, is designed to unveil the "black box" of the model. In the process of data preprocessing, relevant domain knowledge of CFST is harnessed to create new features. Subsequently, feature selection is performed using the Pearson correlation coefficient (PCC), XGBoost, and Random tree algorithms, followed by anomaly detection conducted through the Isolation Tree algorithm. Afterward, the proposed model integrates domain knowledge constraints into the loss function, considering monotonic relationships and approximate constraints between different input parameters and output results. Multiple domain constraints are combined during the model construction to enhance its performance. This was further validated by conducting an ablation test to confirm the superiority of the proposed model and understand the contribution of each loss constraint, providing deeper insights into the model's behavior. The model's robustness is examined to test its stability in extremely noisy environments. Additionally, sensitivity analysis and the SHAP method based on Genetic algorithm are introduced as an interpretability method for the proposed model, aiding in a better understanding of the model's decision-making process and prediction outcomes. In summary, the developed DKNN model and the application of SHAP present a compelling methodology for predicting the bearing capacity of CFSTs. This approach not only contributes significantly to the field of CFST research but also offers valuable insights applicable to analogous research areas.

## 2. Methods

### 2.1 Dataset

This study has meticulously constructed a comprehensive dataset, amalgamating data sourced from physical models with empirical data, a strategy crucial in the realm of domain knowledge-driven machine learning. This integration aims to augment both the predictive accuracy and generalization capabilities of the model. The experimental dataset, compiled by Thai et al. in 2020, includes 1305 circular steel concrete specimens, marking a significant contribution to the field [17]. The dataset includes critical input variables such as component diameter (D), steel tube thickness (t), concrete's compressive strength ($f_c'$), steel tube's compressive strength ($f_y$), and component height (L). Additionally, this research includes data from 768 finite element simulations of high-strength, CFST components(Tran [18])and 548 data points from relevant literature, totaling 2621 samples (see Table 1 in Appendix A). The diversity and extensiveness of this dataset are poised to significantly bolster future research, potentially catalyzing breakthroughs in the development and application of CFST members. The distribution of the dataset is shown in Fig. 1 and Fig. 2 in Appendix A.

Table 2 New features and mathematical expression

| | New features | Mathematical expression |
|---|---|---|
| Geometric and Physical Properties | Steel tube section area ($A_s$) | $A_s = \dfrac{\pi[D^2 - (D - 2t)^2]}{4}$ |
| | Concrete section area ($A_c$) | $A_c = \dfrac{\pi(D - 2t)^2}{4}$ |
| | Steel-concrete section area ($A_{sc}$) | $A_{sc} = A_s + A_c$ |
| | Cross-sectional perimeter ($C$) | $c = \pi D$ |
| | Diameter-to-thickness ratio ($D/t$) | $D/t\ ratio = \dfrac{D}{t}$ |
| | Steel tube volume ($V_s$) | $V_s = \dfrac{\pi[D^2 - (D - 2t)^2]}{4} L$ |
| | Concrete volume ($V_c$) | $V_c = \dfrac{\pi(D - 2t)^2}{4} L$ |
| Structural Performance and Capacity Indicators | Constraint efficiency coefficient ($\xi$) | $\xi = \dfrac{A_s f_y}{A_c f_c'}$ |
| | Nominal Axial Strength ($N_{u0}$) | $N_{u0} = A_s f_y + A_c f_c'$ |
| | Steel tube loading capacity ($N_s$) | $N_s = A_s f_y$ |
| | Concrete loading capacity ($N_c$) | $N_c = A_c f_c'$ |
| | Strength enhancement factor (SEF) | $SEF = \dfrac{D}{t(f/235)^{0.5}}$ |
| Material Proportion and Efficiency Parameters | Steel ratio ($\alpha_{sc}$) | $\alpha_{sc} = \dfrac{A_s}{A_c}$ |
| | Slenderness ratio ($\lambda$) | $\lambda = \dfrac{4L}{D}$ |

## 2.2 Feature engineering and data preprocessing

Feature engineering is a crucial step in ML and deep learning, with the primary objective of extracting and refining information from raw data to represent the problem accurately and predict the target. The process encompasses feature selection and construction, utilizing domain knowledge and in-depth data understanding. Feature extraction involves processing raw data to distill information pertinent to the target variable, while feature selection identifies the most relevant features among those extracted. Meanwhile, feature construction aims to unearth hidden relationships between features, expand the features elements, and expand the feature space by inferring or creating additional features [19,20].

### 2.2.1 Feature construction

In this research, feature construction builds upon the original variables, integrating the expertise of domain knowledge to develop new features using pertinent physical models, statistical methods, and other techniques. These new features are enumerated in Table 2. We performed logarithmic transformations on the label data before training to address biases and inconsistent scaling issues in the training and validation data.

### 2.2.2 Feature selection

Feature selection serves the purpose of simplifying the initial feature set with the objective of achieving various optimization goals. It enhances data clarity by eliminating irrelevant or redundant features, reducing information noise, and rendering the model more comprehensible and interpretable. This preliminary step contributes to the reduction of data complexity, thereby making it more manageable for the model. Meanwhile, the feature selection process addresses the issue of multicollinearity, which involves high inter-feature correlation. Multicollinearity can lead to model instability and challenges in converging to reasonable solutions. The fundamental concept behind feature selection is eliminating superfluous information, reducing feature intercorrelation, and extracting key features, thereby creating a more effective, stable, and generalizable model. This equips models to better address the intricate challenges presented in real-world engineering problems [21]. Consequently, this research employs three distinct methods to assess the importance of the 21 candidate features: PCC, SHAP, and Mean Decrease Impurity (MDI). By these methods, the significance of each feature in the dataset can be systematically evaluated.

#### 2.2.2.1 Pearson correlation coefficient

The correlation matrix is obtained by calculating the PCC method between each pair of variables, measuring the strength and direction of the linear relationship between two variables, with values ranging from -1 to 1. The PCC statistical index

can be determined using the following equation [22]:

$$\rho_{x,y} = cov(X,Y)/\sigma_x \sigma_y \tag{1}$$

Where $\rho_{(x,y)}$ is the PCC value among the matrices comprises of the variants X and Y, $cov(X,Y)$ demonstrates the covariance of X and Y, and $\sigma_x$ and $\sigma_x$ are the standard deviation.

## 2.2.2.2 SHAP value based on XGboost algorithm

By leveraging cooperative game theory, SHAP interpreters provide a comprehensive and coherent approach to understanding the contribution of each feature in the complex ML model. The use of an additive explanation model allows for a more interpretable representation of the original model's behavior, providing valuable insights into feature importance and model decision-making processes. The approach strikes a balance between interpretability and accuracy, making it a powerful tool for explaining complex ML models in various domains [23].

The function g(z') is a linear combination of binary variables, expressed as [24]:

$$f(x) \approx g\left(z^{'}\right) = \varphi_0 + \sum_{i=1}^{M} \varphi_0 z_i^{'} \tag{2}$$

where x presents the original input features. The interpretability model uses x' as the simplified input features and links them to the mapping function $x = h(x')$, such that the local approximation $g(z') \approx f(h(x(z')))$ holds whenever $z' \approx x'$. The value φ corresponds to the Shapley value, which is expressed as:

$$\varphi_i(f,x) = \sum_{z^{'} \in x^{'}} \frac{|z^{'}|!(M-|z^{'}|-1)!}{M!} (f_x\left(z^{'}\right) - f_x\left(z^{'}|i\right) \tag{3}$$

where N represents the set of all input features, D denotes the dimensionality of all input features, which represents the feature coalition with nonzero indices related to the existing features in z'. Additionally, it corresponds to the expectation of the function conditioned on the subset S of input features:

$$C_i = \frac{1}{n_{total}} \sum_{j=1}^{n_{total}} |\varphi_{ij}| \tag{4}$$

where the symbol $n_{total}$ represents the total number of samples in the dataset, the term $\varphi_{ij}$ denotes the contribution of feature i to sample j, and $C_i$ represents the global contribution of feature i.

In this section, the SHAP values are computed using the XGBoost algorithm, a tree-based ensemble learning model composed of multiple decision trees. The underlying principle of XGBoost involves iteratively traversing all paths in the decision trees and calculating the contribution of each feature to the prediction. The pseudocode is outlined in Table 2 in Appendix A:

## 2.2.2.3 MDI value based on Random Forest algorithm

In Random Forest, MDI is utilized to assess the importance of each feature by calculating the decrease in Gini impurity resulting from splitting the dataset based on that feature in each decision tree. As illustrated in Table 3 in Appendix A, every feature in every decision tree is used to split the dataset, and the difference in Gini impurity before and after the split is computed. Then, the average of these Gini impurity reductions across all trees is taken to obtain the MDI importance score for that feature. MDI serves as a metric to measure the significance of features in Random Forest, providing valuable insights into their contribution to the predictive performance [25].

The formula for calculating Gini impurity is as follows [26]:

$$Gini(D) = 1 - \sum_{k=1}^{k}(p_k)^2 \qquad (5)$$

where $D$ represents the dataset, $k$ denotes the number of categories in the target variable, and $p_k$ corresponds to the probability of the target variable belonging to the $k_{th}$ category.

## 2.2.3 Isolation Forest-Based Anomaly Detection

This section aims to detect anomaly detection on the experimental dataset of CFST component under axial loading. The goal is to identify potential outlier samples, assisting engineers and researchers in accurately assessing the reliability and safety of structures. CFST structures find wide applications in engineering, and their performance evaluation relies heavily on abundant experimental data. However, these datasets may contain certain levels of anomalies due to experimental errors, equipment malfunctions, or other external factors. If these outliers are not identified and addressed promptly, they may lead to potential structural design or construction issues. In this study, the Isolation Forest algorithm is adopted as the method for anomaly detection. This algorithm excels at isolating abnormal instances by constructing random forests and isolating them in shorter branches. By leveraging this approach, it becomes possible to effectively detect and handle outliers within the dataset.

The Isolation Forest algorithm is an unsupervised ensemble learning method that aims to detect anomalies by isolating them in the feature space by constructing random decision trees. We set an appropriate proportion of anomaly points to control the detection sensitivity of the Isolation Forest algorithm [27]. After the training, we utilize the trained model to predict the samples in the dataset and identify potential outlier instances. The algorithmic operation process is presented in Table 4 in Appendix A.

The specific formula for calculating the anomaly score is as follows [28]:

$$S(x,n) = 2^{-\frac{E(h(x))}{c(n)}} \qquad (6)$$

where *x* represents the sample point, n denotes the size of the training dataset, *h(x)* corresponds to the depth (height) of the sample point *x* within an individual Isolation Tree, *E(h(x))* signifies the average depth of sample point x across all randomly constructed Isolation Trees. *c(n)* represents the expected value of the average path length and can be calculated.

## 2.3 DKNN models

The structural framework of the Domain knowledge-enhanced Neural Network (DKNN) model is illustrated in Fig. 1(a). Traditional ML models typically employ a loss function comprising two primary components: a supervisory signal derived from labelled data and a regularization element to avert overfitting. However, in knowledge-based terms, it is integrated into the loss function. The research introduces domain knowledge into deep NNs through a penalty term, enforcing constraints reflective of this knowledge. The optimization process during training minimizes the total loss, which includes this penalty term, thus ensuring that domain-specific restrictions are applied to the model's predictions and output range. Domain knowledge is further encoded as logical constraints, each transformed into a loss term [10]. The hybrid loss function for these models is expressed as:

$$Loss = Loss_{train}(Y_{true}, Y_{pred}) + \gamma Loss_{phy}(Y_{pred}) \quad (7a)$$

$$Loss_{phy}(Y_{pred}) = Loss_D(Y_{pred}) + Loss_D(Y_{pred1}, Y_{pred2}) \quad (7b)$$

Here, $Loss_{train}(Y_{true}, Y_{pred})$ represents the conventional standard training loss, where $Y_{true}$, and $Y_{pred}$ signify the ground truth and predicted values, respectively. $Loss_{phy}(Y_{pred})$ constitutes the domain-specific loss, with $\gamma$ weighting the domain knowledge constraints.

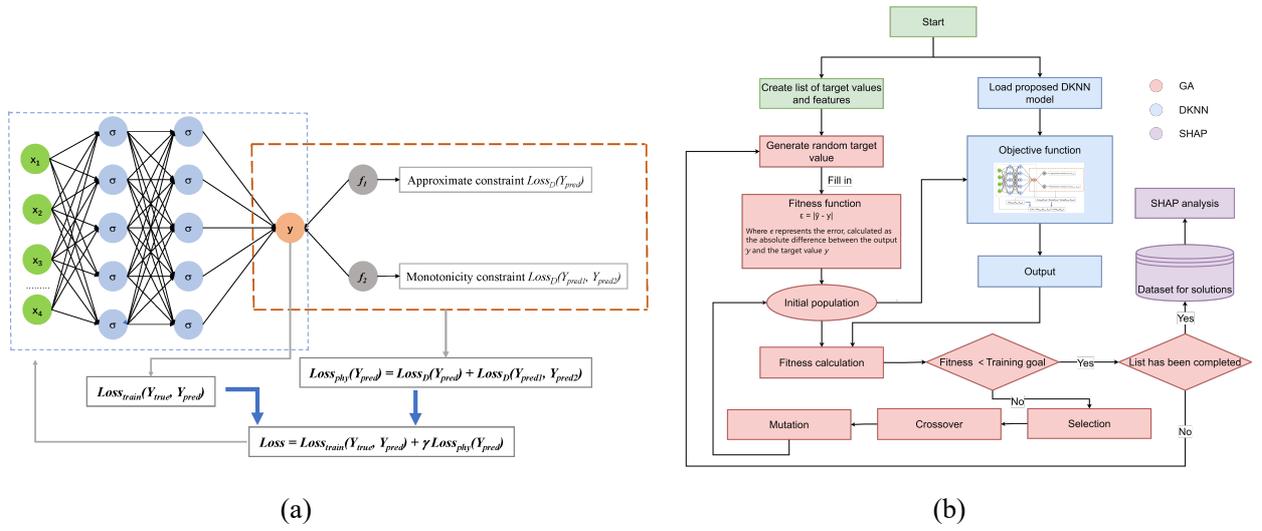

(a)                  (b)

Fig. 1. Structure of DKNN model and flowchart of SHAP analysis

## 2.3.1 Approximate constraints

In practical scenarios, measurements may deviate, leading to noisy datasets that could diminish model accuracy. Domain knowledge insights on the feasible operational range of the target variable can guide the training of more precise models. These insights inform the specification of approximate constraints for the target variable Y, denoted as $[y_l, y_u]$. These constraints are integrated into the training loss function of the neural network as follows: [9].

$$g(Y_{pred}) = f(x) = \begin{cases} 0, & if\ Y_{pred} \in [y_l, y_u] \\ |y_l - Y_{pred}|, & if\ Y_{pred} < y_l \\ |y_u - Y_{pred}|, & if\ Y_{pred} > y_u \end{cases} \quad (8a)$$

$$Loss_D(Y_{pred}) = \sum_{i=1}^m ReLU(y_l - y^i) - ReLU(y^i - y_u) \quad (8b)$$

$$ReLU(z) = z^+ = max\,(0, z) \quad (8c)$$

## 2.3.2 Monotonicity Constraint

Specific engineering parameters often exhibit a monotonic relationship. The correlation heatmap demonstrates that parameters such as $A_s$, $A_c$, $A_{sc}$, $D$, $C$, $N_{u0}$, $N_s$, $V_s$, $V_c$, and $N$ share a robust linear relationship, with PCC values exceeding 0.7. This indicates a strong linear relationship between these input and output parameters $N$. The mechanics performance of CFST columns is based on composite material theories, wherein steel tubes and core concrete are combined to withstand both pressure and external loads collectively. Therefore, the augmentation of features related to geometric parameters, such as $D$, $C$, $A_s$, $A_c$, $A_{sc}$, $V_s$, and $V_c$, effectively results in the increased utilization of materials, thereby enhancing the overall load-bearing capacity of the structure. This enhancement occurs because these parameters directly influence the geometric characteristics of the composite material, such as cross-sectional area, diameter, and perimeter, subsequently elevating the column's overall stability and load-bearing capacity. The values of $N_s$ and $N_{u0}$ are characterized by their sensitivity not only to geometric parameters but also to the distinct influence of material properties. Specifically, they take into consideration the properties of two crucial materials, namely the yield strength ($f_y$) of steel tubes and the compressive strength ($f_c'$) of concrete. $N_s$ represents the contribution of the steel tubes in the composite structure. Its increase has a positive impact on the overall load-bearing capacity of the composite material due to the provision of additional strength. This extra strength enables the steel tubes to better withstand pressure, consequently enhancing the stability and load-bearing capacity of the overall structure. On the other hand, $N_{u0}$ accounts for the contributions of both steel tubes and concrete, collectively enhancing the overall load-bearing capacity. By comprehensively considering the properties of both steel tubes and concrete, the increase in $N_{u0}$ effectively raises the structural load-bearing capacity. This comprehensive approach in considering material properties contributes to a more comprehensive understanding of the load-bearing capacity of composite structures. Therefore, a monotonic increasing relationship can be established among $A_s$, $A_c$, $A_{sc}$, $D$, $C$, $N_{u0}$, $N_s$, $V_s$, $V_c$, and $N$. The mathematical expression for this can be described as follows:

$$Loss_D(Y_{pred1}, Y_{pred2}) = \sum_{i=1}^{m} IdX((x_1^i < x_2^i) \wedge (y_{pred1}^i > y_{pred2}^i)) \cdot ReLU(y_{pred1}^i - y_{pred2}^i) + \sum_{i=1}^{m} IdX((x_1^i < x_2^i) \wedge$$

$$(y^i_{pred1} > y^i_{pred2})) \cdot ReLU(y^i_{pred2} - y^i_{pred1}) \qquad (9)$$

Where $x_1$ and $x_2$ represent the measured values of individual phenomena in various contexts within the system ($x_1$ and $x_2$ could represent different material strengths, thicknesses of steel tubes, and component areas, etc.), *IdX(·)* represents the identity function that evaluates to true if the result of the logical AND ($\wedge$) operation evaluates to true and is false otherwise. The identity function *IdX* generates a boolean mask to identify instances where measurements obey the monotonicity constraint being enforced while the predictions by the neural network model violate the constraint. Applying this mask to the *ReLU* function allows us to capture errors only of the instances wherein the domain constraint is violated. Formulating the domain loss *LossD (·)* in this manner causes the model to change course to a region in the (learned) function space more amenable to the injected domain constraint.

### 2.3.3 Genetic algorithm based SHAP analysis

We adopted a comprehensive approach, combining the proposed DKNN model, Genetic Algorithm (GA), and SHAP value analysis to thoroughly investigate the impact of two key features ($f_c'$ and $\alpha_{sc}$) on the model output. Based on the proposed prediction model, we introduced Genetic Algorithm (GA) for inverse problem-solving. The objective of this process is to reverse-engineer the model output to closely match predefined target values for specific $f_c'$ and $\alpha_{sc}$ feature values. To achieve this goal, we introduced the calculation of each solution's fitness by using the set target values. Specifically, the GA was employed to search for a batch of sample points covering various combinations of the $f_c'$ and $\alpha_{sc}$ features under fixed target values. This optimization process involved adjusting the input parameters of the proposed DKNN model to design control variables. The flowchart is presented in Fig. 1(b).

A total of 480 samples were created and then subjected to SHAP value analysis. Through controlled variable experiments on these sample points, we validated the model's interpretability and generalization performance concerning these two features. The application of this integrated approach provides a comprehensive and in-depth perspective, enabling a better understanding of the roles of these features in the model, and offers robust support for future material design and research.

## 2.3.4 Evaluation matrix

This study utilizes four principal metrics to assess the regression model's performance, as detailed in Table 3. These metrics are organized into three main evaluation categories: accuracy, variability, and overall effectiveness. In the accuracy domain, the study calculates both Root Mean Squared Error (RMSE) and Mean Absolute Percentage Error (MAPE). MAPE evaluates the relative accuracy of predictions, while RMSE measures error magnitude, offering a dual perspective on error analysis for a nuanced understanding of model accuracy; For overall model effectiveness, the research applies the Coefficient of Determination ($R^2$), quantifying how well the model explains data variability. Meanwhile, the Coefficient of Variation (CoV) [1,29,30] assesses the model's consistency or prediction dispersion. By classifying these metrics into categories based on their evaluative focus, the research achieves a comprehensive analysis of the regression model's performance across multiple dimensions.

Table 3 Evaluation Metrics

| Metrics | Mathematical expressions | Definitions |
|---|---|---|
| Root Mean Squared Error | $RMSE = \sqrt{\frac{1}{N}\sum_{i=1}^{N}(t_i-\alpha_i)^2}$ | The square root of the MSE, indicating the average magnitude of error without considering direction |
| Mean Absolute Percentage Error | $MAPE = \frac{1}{N}\sum_{i=1}^{N}\left|\frac{t_i-\alpha_i}{\alpha_i}\right|$ | The average percentage difference between predicted and actual values relative to actual values. |
| Coefficient of Determination | $R^2 = \frac{\sum_{i=1}^{N}(t_i-\alpha_i)^2}{\sum_{i=1}^{N}(\bar{t}_i-\alpha_i)^2}$ | A measure indicating the proportion of variance in the dependent variable explained by the independent variables |
| Coefficient of Variation | $CoV = \frac{Bias - (t_i-\alpha_i)}{Bias}$ | The ratio of the standard deviation to the mean, representing the variability relative to the mean value |

Explanation: Where $N$ denotes the number of examples $t_i$, and $\alpha_i$ are the target and predicted values of the $i^{th}$ sample.

## 2.3.5 Robustness evaluation

To evaluate model robustness, the perturbation was applied exclusively to the training labels, denoted as $y_{train}$. The process is governed by two key parameters: the perturbation sample ratio $p$ and the perturbation range $d$. The former dictates the proportion of the dataset to be perturbed, while the latter defines the magnitude of the perturbation. The mathematical expression for the perturbed label, $y'$, is given by:

$$y' = \begin{cases} y \cdot (1 + d), & if\ |d| \leq p \\ y, & otherwise \end{cases} \quad (10)$$

where $y$ is the original label and $d$ is a random number generator producing values uniformly distributed in the range [-1, 1]. During the training phase, each label in the dataset is evaluated independently. A random number is generated for each label, and if this number falls below or equals to the perturbation magnitude threshold $p$, the label is modified according to the formula for $y'$. This selective perturbation introduces variability into the training process, aiming to enhance the model's robustness and generalization capabilities.

The perturbed dataset is then employed to train the model. Its performance is benchmarked against a model trained on an unperturbed dataset, facilitating a comparative analysis to ascertain the influence of data perturbation on the model's generalization from noisy or varied input data.

## 3. Result and Discussion
### 3.1 Feature engineering and data preprocessing results

This section of the study presents a comparative analysis of feature selection importance rankings utilizing three distinct methodologies: PCC, SHAP values, and MDI. The comparative results are graphically represented in Fig. 2. An examination of the outcomes from these methods reveals a degree of congruence. As detailed in Table 5 in Appendix A, features such as $N_{u0}$, $A_s$, $V_c$, $V_s$, and $D$ consistently appear within the top ten across all three methodologies. Furthermore, features $A_c$, $A_{sc}$, $C$, and $N_s$ are identified as significant in both the MDI and PCC methods. These features exhibit strong linear relationships with the outcome, as indicated by high absolute correlation coefficients in the Pearson correlation method. Additionally, their contributions in reducing impurity in the MDI method are noteworthy. The feature $f_c'$ also emerges in the top ten rankings for both the MDI and SHAP methodologies.

Nevertheless, it is critical to acknowledge the inconsistencies across the results derived from the correlation heatmap, MDI, and SHAP values. These discrepancies stem from the varying metrics and computational approaches employed by each method. While the correlation heatmap focuses primarily on linear relationships, SHAP values and MDI encompass more intricate models and data distributions. Consequently, the correlation heatmap might not fully elucidate the significance of features when nonlinear relationships are present between features and the target variable.

To circumvent the disparities among the three methods, an integrated analysis of their results facilitates a more robust feature selection approach. Consequently, the following ten features have been selected as input parameters for the model: $N_{u0}$, $A_s$, $V_c$, $V_s$, $D$, $A_c$, $A_{sc}$, $C$, $N_s$, and $f_c'$. This selection strategy aims to simplify model complexity while enhancing prediction accuracy and generalization capabilities. It also provides a more comprehensive understanding and prediction of the bearing capacity of CFSTs.

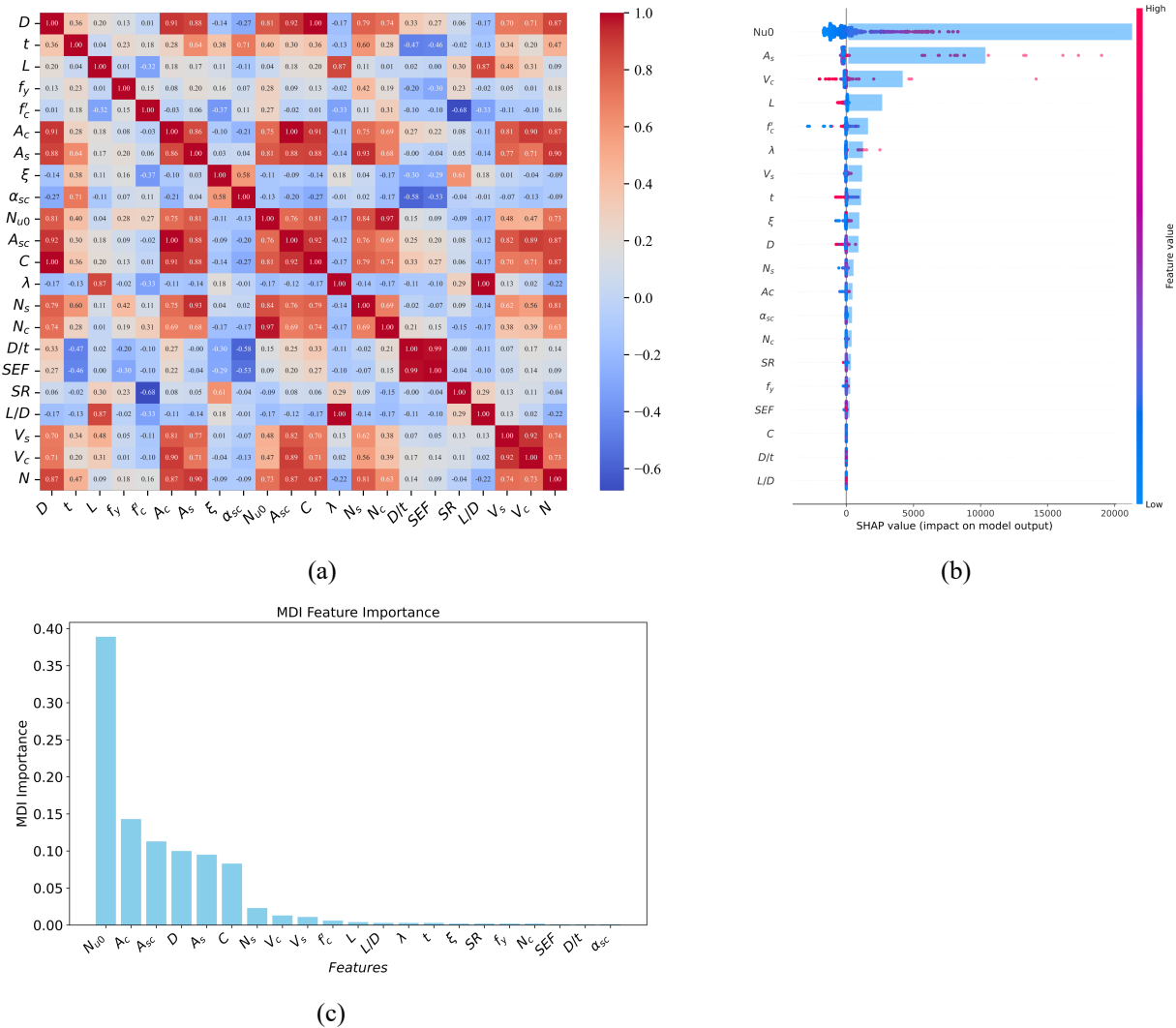

Fig. 2. Comparison of feature contributions ranking using (a) PCC, (b) SHAP methods and (c) MDI

The newly created features ($A_s$, $A_c$, $A_{sc}$, $C$, $N_{u0}$, $N_s$, $V_s$, $V_c$, and $N_c$) demonstrate higher PCC values in relation to the axial compressive bearing capacity of CFSTs, indicating that they capture stronger linear relationships with the prediction outcomes compared to the original input parameters. Moreover, according to SHAP value analysis, features $N_{u0}$, $A_s$, and $V_c$ ranked among the top three in importance, underscoring the value of feature construction in amplifying the representational capacity of underlying patterns and relationship in the data.

Incorporating domain knowledge into feature engineering aims to capture critical aspects and characteristics of CESTs that maybe overlooked by original input features. For instance, the new feature $N_{u0}$ represents a normalized measure related to the first mode shape of the structure, encapsulating its fundamental dynamic behavior, which is vital for understanding structural stability and bearing capacity under various loads. Similarly, features $A_s$, $A_c$, $A_{sc}$, $C$, $N_s$, $V_s$, $V_c$, and $N_c$, derived from geometric and mechanical properties of the CFSTs, play a significant role in defining the overall structural behavior. The introduction of these domain knowledge-based features allows the model to interpret more relevant data, thereby enhancing predictive accuracy.

## 3.2 DKNN performance assessment
### 3.2.1 Model development

This study conducted an extensive examination of neural network models with varying hidden layers numbers for predicting the bearing capacity of CFSTs. The models evaluated include the baseline model (ANN), the model with approximate constraints (ANNWA), the model with monotonic constraints (ANNWM), and the model with integrated constraints (ANNWT).

As depicted in Fig. 3(a), the plot provides insights into the effect of hidden layers on model performance, using metrics such as MAPE and RMSE. For the ANN, ANNWM, and ANNWT models, an increase in hidden layers correlates with a decrease in those evaluation metric values. This trend suggests improved model performance, characterized by reduced percentage errors, and a heightened ability to discern complex data patterns, thereby elevating prediction accuracy for component bearing capacity. Mainly, the progression of hidden layers signifies a more intricate functional form, facilitating a more precise fitting of nonlinear relationships within the data. Consequently, NNs can adeptly adjust to progressively complex data patterns. Furthermore, networks with multiple hidden layers can perform hierarchical feature extraction. Each layer can capture distinct levels of abstraction within the data, and the amalgamation of these features enables a more comprehensive depiction of data complexity [31]. Conversely, the ANNWA model exhibited a more nuanced trend: an initial increase in evaluation metrics, followed by a sharp decline and a subsequent minor rise. Despite an overall trend of decreasing evaluation metrics with additional hidden layers, some nonlinear characteristics were observed, particularly with small increases in hidden layers. This is due to the increase in model complexity, which may lead to overfitting to the training data and consequently result in a decline in performance on the test dataset [32].

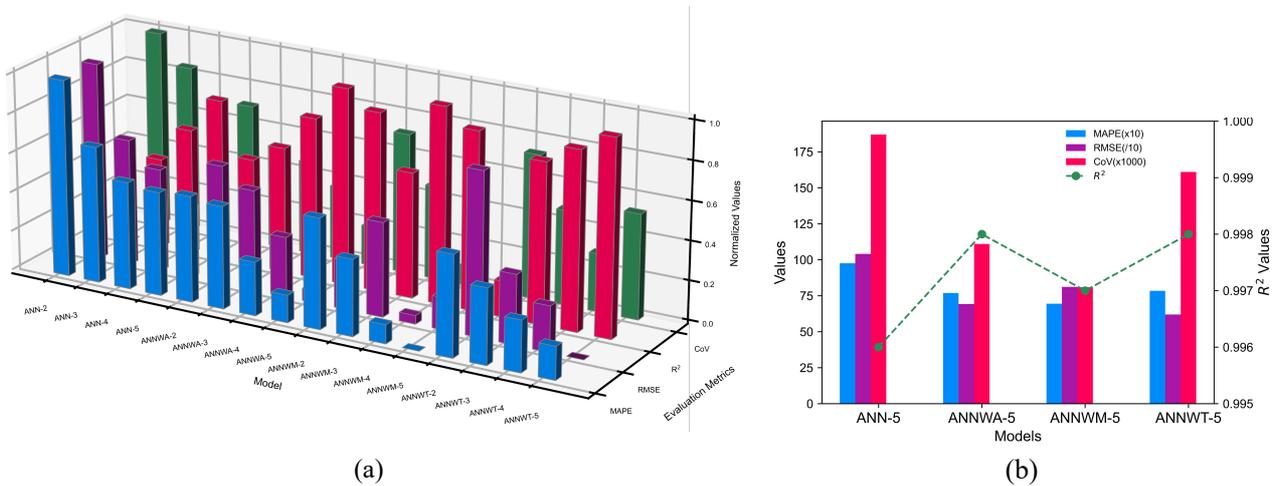

Fig. 3. (a) Performance assessment of the models and (b) comparing with the benchmark model's performance

Model ANNWT-5 exhibits significantly lower prediction errors compared to other models. It presents superior predictive accuracy, with an MSE of 383,787.944, MAE of 318.371, and RMSE of 619.506, as shown in Table 6 in Appendix A. Regarding MAPE, model ANNWM-4 achieves the lowest value at 6.944%. Despite a slightly higher MAPE of 7.844%, ANNWT-5 still demonstrates exemplary performance among all models. ANNWT-5 demonstrates low levels of both overall and absolute error, highlighting its exceptional performance in precisely representing observed data.

Further, ANNWT-5 has a good performance of fit between the predicted results and actual observations as well as presents a low dispersion. Fig. 3(a) reveals that the R2 value for ANNWT-5 reaches 0.998, paralleling the top performance of models ANNWA-5 and ANNWM-4, substantiating the strong correlation between model predictions and experimental data. Additionally, the value of CoV is 0.161, indicating that most model predictions closely align with the actual observed values. A lower value of CoV indicates a reduction of variation of data points and an increase of consistency and stability within the data, exhibiting smaller fluctuations. These results highlight that ANNWT-5, with its dual constraints, excels in various performance evaluations, showing high predictive accuracy and close alignment with actual observed data.

The study then delves into the impact of domain knowledge constraints on neural network model performance. The objective is to ascertain how these constraints influence model efficacy and to emphasize the integral role of combining data-driven approaches with theoretical knowledge. A comparative analysis between models with constraints and the conventional ANN model, ANN-5, was undertaken (see Fig. 4(b)). Models with constraints, whether single or dual, exhibit marked performance improvements. Model ANNWT-5, which integrates two constraints, shows the most pronounced enhancement. Its reductions of RMSE reaches 40.470%. The ANNWA-5 model, incorporating an approximate constraint, also demonstrates significant performance gains, with decreases of 33.474% in RMSE. While the improvements for model ANNWM-5, with a monotonicity constraint, are comparatively modest, it still shows a reduction of 22.001%. The improvement in $R^2$ compared to other metrics is not as pronounced. This is due to the benchmark model ANN-5 already having an $R^2$ value very close to 1. Although the increase of 0.002 seems small in absolute terms, within the high numerical range near the extreme value of 1, this slight change could actually have a significant substantive impact.

These findings vividly illustrate that the introduction of constraints significantly bolsters model performance, reducing errors and enhancing prediction accuracy. The study unequivocally confirms the substantial benefits of introducing domain knowledge constraints in improving model performance, particularly as evidenced by the exceptional performance of the ANNWT model across all metrics. This finding highlights the crucial role of harmonizing data and theoretical knowledge in advancing machine learning capabilities, leading to more accurate predictions and superior performance.

### 3.2.2 Robustness analysis of model

The robustness of the models, ANN-5 and ANNWT-5, was evaluated under varying noise levels, as depicted in Fig. 4(a) and outlined in Table 4. Both models demonstrate an uptrend in performance with increasing noise percentages, particularly around ρ=30%. This trend is more pronounced in the ANN-5 model, whereas the ANNWT-5 model shows a diminishing rate of performance degradation, stabilizing post ρ=30%. Remarkably, even at extreme noise levels approaching ρ=50%, ANNWT-5's performance substantially outperforms that of ANN-5. Quantitatively, the values of MAPE for ANN-5 and ANNWT-5 increases by 84.488% and 76.492%, respectively, as noise levels escalate from 10% to 50%. Notably, the rise in MAPE for ANNWT-5 is less steep compared to ANN-5, indicating better noise resilience.

In environments with high noise, establishing a minimum acceptable error threshold is a standard approach. Thus, identifying robust function space representation methods is crucial. For example, if a MAPE of 15% is acceptable, the DKNN model maintains this threshold even at a noise level of approximately 50%, while the Neural Network (NN) model exceeds it at 30% noise.

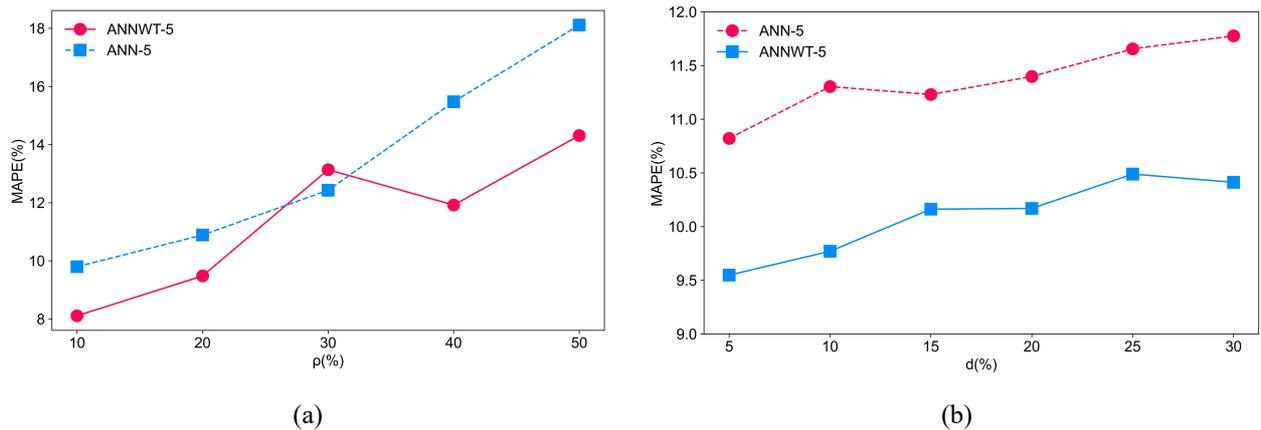

(a)                                (b)

Fig. 4. Noisy validation

Table 4 Noise percentage vs. MAPE

| Model types | $\rho$ | | | | |
|---|---|---|---|---|---|
| | 10% | 20% | 30% | 40% | 50% |
| ANNWT-5 | 8.108 | 9.48 | 13.135 | 11.919 | 14.31 |
| ANN-5 | 9.797 | 10.887 | 12.427 | 15.472 | 18.113 |

Table 5 Error magnitude vs. MAPE

| Model types | $d$ | | | | | |
|---|---|---|---|---|---|---|
| | 5% | 10% | 15% | 20% | 25% | 30% |
| ANNWT-5 | 9.548 | 9.77 | 10.162 | 10.169 | 10.489 | 10.412 |
| ANN-5 | 10.821 | 11.304 | 11.23 | 11.398 | 11.655 | 11.776 |

When varying the error magnitude (d), bothANN-5 and ANNWT-5 models showed a consistent increase in MAPE as observed in Fig.4(b) and Table 5. The change in MAPE for ANN-5 was more pronounced, increasing by 0.955%. Conversely, ANNWT-5 exhibited a smaller increase by 0.864%. This trend suggests that the ANN-5 model is more stable than ANNWT-5 in response to perturbations.

Despite the continuous increase in error magnitude, ANNWT-5 consistently exhibited a lower MAPE compared to ANN-5. Notably, at a 30% error magnitude, ANNWT-5's MAPE remained lower than ANN-5's MAPE at a mere 5% error. These findings highlight the superiority of domain-knowledge-enhanced ANN models in handling significant training data errors over models trained without domain knowledge but with smaller training data errors.

In conclusion, under high-noise conditions, the ANNWT-5 model demonstrates superior stability and robustness compared to the ANN-5 model. The ANN-5 model is more susceptible to noise, particularly in domain-agnostic scenarios, while ANNWT-5 effectively mitigates noise interference, maintaining stable performance levels. Amid external disturbances, ANNWT-5 adapts more effectively to data variations, sustaining its predictive capabilities.

### 3.2.3 Model ANNWT-5 performance assessment

The ANNWT-5 model, with five hidden layers, was selected for detailed performance analysis. Fig. 5(a) reveals a linear relationship between the model's predictions and actual observations, with data points clustering around the 45-degree diagonal line. This indicates the ANNWT-5 model's proficiency in capturing linear relationships and fitting the data accurately. The model achieved prediction errors below+10% for 75.619% of the data points and maintained errors below +20% for 91.810%of them.

Fig. 5(b) displays the residuals for each prediction. Although some data points showed relatively high residuals, these were generally linked to larger observed value sand the discrepancies between actual and predicted values for these high-residual points remained within 20%. It indicates that the overall model performance remained robust. Table 6 presents the MAPE performance across different concrete and steel tube strength intervals. This detailed analysis of MAPE values allows for a comprehensive assessment of the model's performance under various strength conditions.

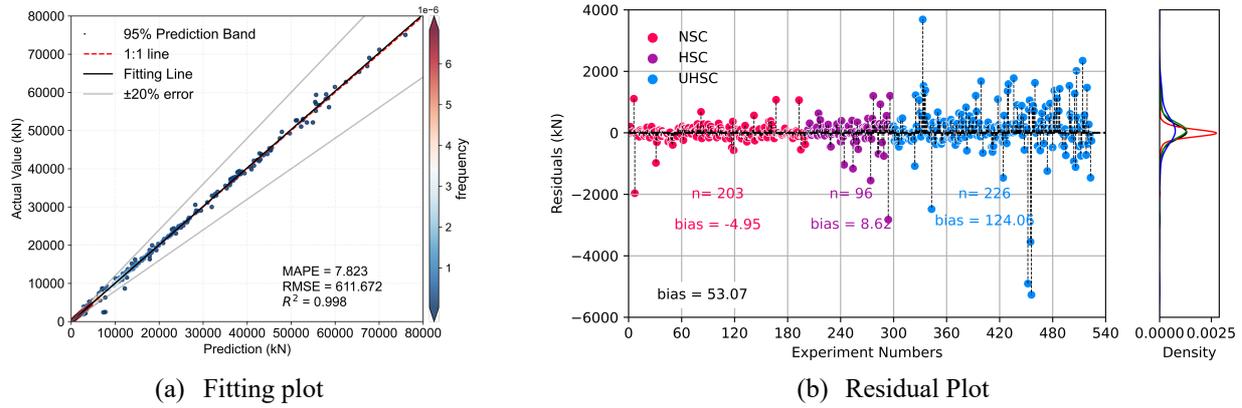

(a) Fitting plot  (b) Residual Plot

Fig. 5. Comparison between the predicted values and the experimental values from the validation dataset.

Table 6 Model performance across different concrete and steel tube strength intervals

|  |  | MAPE (%) |  | $f_c'$ (MPa) |  |  |
|---|---|---|---|---|---|---|
|  |  | NSC | HSC | UHSC | Total |  |
| $f_y$ (MPa) | NSS | 8.949 | 6.171 | 7.774 | 7.956 |  |
|  | HSS | 12.829 | 5.392 | 5.291 | 7.238 |  |
|  | Total | 9.350 | 6.066 | 7.246 | 7.844 |  |

Overall, MAPE values for all the strength intervals were below 13% and the average MAPE for the entire dataset was observed to be 7.844%, underscoring the model's consistently high predictive accuracy across intervals. However, performance variations were noted within specific strength intervals. CFST combinations of High-Strength Steel (HSS) with Ultra-High Strength Concrete (UHSC) exhibited the best performance, with a MAPE of 5.291%. Conversely, the combination of HSS with Normal Strength Concrete (NSC) had the highest MAPE at 12.829%. This is because there are fewer combinations of NSC and HSS in practical engineering, resulting in a lower representation of such components in the entire dataset. Additionally, due to the relatively small proportion of samples representing Ultra High Strength Steel (UHSS) components in the entire database, and the lack of actual sample analysis in combination with UHSC, there will be no independent discussion on this type of components. Furthermore, while the MAPE differences between different steel tube strength intervals were minimal, the differences between concrete strength intervals were more pronounced. This indicates that the model's performance was relatively balanced in adapting to varying steel tube strengths but exhibited more significant variations with different concrete strengths. This variation likely due to variations in material properties and structural designs.

### 3.2.4 Comparison with design codes and existing models

Table 7 Strength prediction formulas of circular CFST columns.

| Design codes | Expressions | Refs |
|---|---|---|
| AIJ | $P_{AIJ} = 1.27 A_s f_y + A_c f_c'$ | [33] |
| EC4 | $P_{EC4} = \eta_s A_s f_y + \eta_c A_c f_c'$ | [34] |
|  | $\eta_s = 0.25(3 + 2\lambda)$ |  |
|  | $\eta_c = 4.9 - 18.5\lambda + 17\lambda^2$ |  |
| ACI | $P_{ACI} = A_s f_y + 0.85 A_c f_c'$ | [35] |
| GB90536 | $P_{GB} = 0.9 A_c f_{ck}(1 + \theta + \sqrt{\theta})$ | [36] |
| GEP | $P_{GEP} = A_s + 2f_c' - 4\lambda + \sqrt{f_c'}\left(A_c + \sqrt{3f_c' - 9.596}\right) + \dfrac{0.169 A_s (f_y - 2\lambda)\sqrt{A_c - 11.562}}{\left(\dfrac{D}{t}\right)}$ | [37] |
| Han | $P_H = (1.14 + 1.02\theta) f_{ck}(A_s + A_c)$ | [38] |
| Wan | $P_w = \eta_a A_s f_y + \eta_c A_c f_c'$ | [39] |
|  | $\eta_a = 0.95 - 12.6 f_y^{-0.85} \ln\left(\dfrac{0.14 D}{t}\right)$ |  |
|  | $\eta_c = 0.99 + [5.04 - 2.37\left(\dfrac{D}{t}\right)^{0.04}\left(f_c'\right)^{0.1}]\left(\dfrac{t f_y}{D f_c'}\right)^{0.51}$ |  |
| ANN | - | [40] |
| Explanations | $A_s$ and $A_c$ : the cross-sectional area of the steel tubular and the concrete core |  |
|  | $f_y$: the yield strength of the steel tubular |  |
|  | $f_c'$ and $f_{ck}$: the cylinder and cubic compressive strength of the concrete core |  |
|  | $\lambda$: the relative slenderness of CSFTs |  |
|  | $\theta$: the confine coefficient of CFSTs |  |

Table 7 presents a statistical comparison between the proposedANNWT-5 model and estimates from design codes and analytical models found in the literature. It is crucial to acknowledge the imitations of some codes, especially in terms of concrete's compressive capacity, which may limit the applicability of their formulas to all database specimens.

Table 8 illustrates that the ANNWT-5model significantly outperforms other design codes and models. The model's $R^2$ value exceeds 0.98, surpassing the existing Wan's model's highest $R^2$ value of 0.966. Additionally, the ANNWT-5 model's MAPE of 7.844% significantly outperforms the best literature model ANN which has a MAPE of 16.757%. This indicates a reduction of over 50% compared to existing methodologies. Moreover, the ANNWT-5 model's CoV is 0.161. For a granular understanding of error metrics in codes and analytical models, Fig. 6. visually portrays the MAPE, RMSE, $R^2$ and CoV values for the database specimens. These visual representations provide a clearer perspective on the performance variations across different methodologies, bolstering the conclusions derived from the tabulated data.

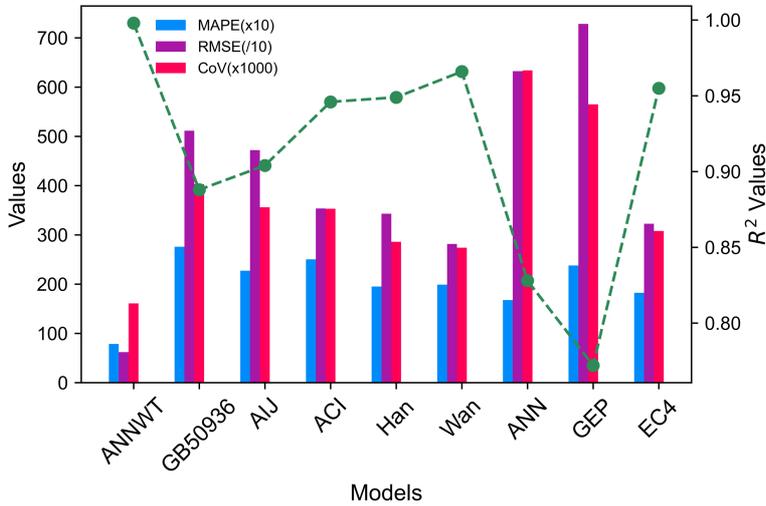

Fig. 6. Error indices for circular CFST columns

Table 8 Statistical assessment and comparison of the proposed model with code-specified formulae, analytical formulae, and machine learning models in the literature

| Formula | MAPE | $R^2$ | RMSE | CoV |
| --- | --- | --- | --- | --- |
| ANNWT | 7.844 | 0.998 | 619.506 | 0.161 |
| GB 50936-2014 | 27.589 | 0.888 | 5112.754 | 0.403 |
| AIJ | 22.711 | 0.904 | 4719.149 | 0.356 |
| ACI | 25.032 | 0.946 | 3534.481 | 0.353 |
| Han | 19.521 | 0.949 | 3428.975 | 0.286 |
| Wan | 19.884 | 0.966 | 2816.099 | 0.274 |
| ANN | 16.757 | 0.828 | 6324.444 | 0.634 |
| GEP | 23.774 | 0.772 | 7282.201 | 0.565 |
| EC4 | 18.244 | 0.955 | 3224.893 | 0.308 |

## 3.3 Parametric analysis and design guidance
### 3.3.1 Sensitivity Analysis

Sensitivity analysis elucidates how variations in input variables influence the dependent variable. In this study, parametric studies and sensitivity analyses were conducted to assess the influence of different input variables on predicting the ultimate load-bearing capacity of CFST columns. The method, following Gandomi et al.'s model [37], measures the impact of each

input variable on the model's output:

$$\text{Sensitivity value} = \frac{f_{max}(x_i) - f_{min}(x_i)}{\sum_{k=1}^{n}(f_{max}(x_k) - f_{min}(x_k))} \times 100 \quad (11)$$

where $f_{max}(x_i)$ and $f_{min}(x_i)$ represents the maximum and minimum prediction values of the model for the range of i-th input variable, while other variables are kept constant at their mean values.

Sensitivity analysis results for CFSTs, as depicted Fig. 7(a), reveal that in circular section columns, two key parameters notably impact the outcomes. Specifically, the volume of the concrete ($V_c$) and the circumference of the column (C) significantly influence the results, contributing approximately 21.7% and 19.4% respectively. Other influential factors include the volume of the steel tube ($V_s$) and the diameter of the cross-section (D), contributing 15.0% and 13.3% to result variations. The area of the concrete ($A_c$) also accounts for a notable 11.8%. Collectively, these five parameters contribute over 80% to the output variations.

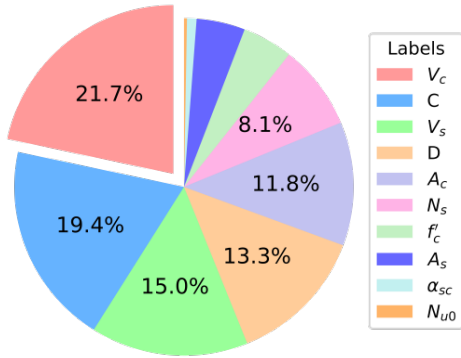

(a) Sensitivity analysis

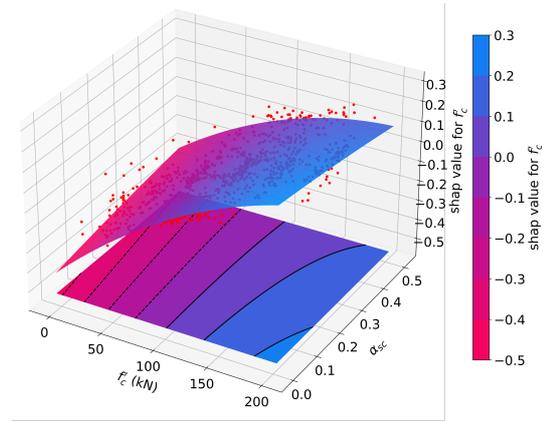

(b) SHAP analysis for $f_c'$

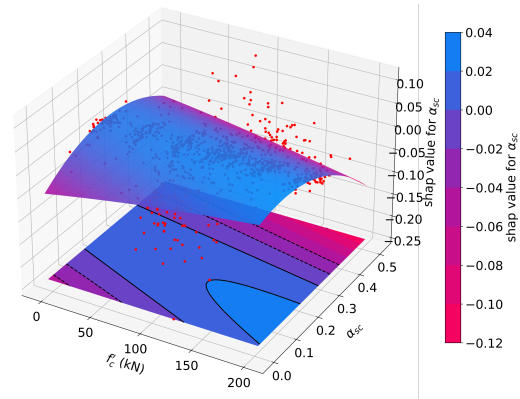

(c) SHAP analysis for $\alpha_{sc}$

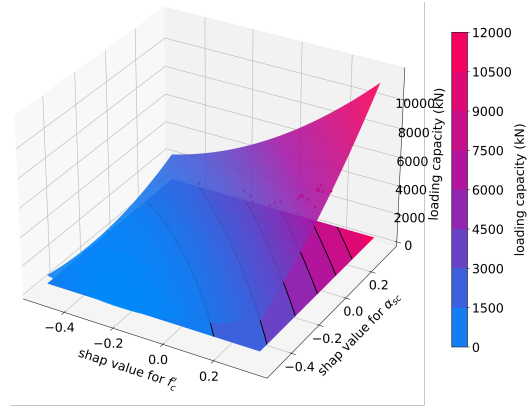

(d) SHAP analysis for $\alpha_{sc}$ and $f_c'$

Fig. 7. Sensitivity analysis and dependence plot

## 3.3.2 Design guidance based on SHAP analysis

By observing the surface plots in Fig. 7(b), a clear nonlinear relationship between concrete compressive strength and steel ratio is identified. In the range of $α_{sc}$ from 0.1 to 0.5, the Shap value for $f_c'$ shows an increasing trend with the rise of $f_c'$, indicating a significant positive impact of $f_c'$ on the load-bearing capacity of CFST members. As $f_c'$ increases, the upward trend in SHAP value for $f_c'$ gradually diminishes, suggesting a reduction in the positive influence of increasing $f_c'$ on the load-bearing capacity. This could be attributed to the saturation of CSFT structures at a certain strength, resulting in a diminishing effect of increasing concrete strength on load-bearing capacity. With the increase of $α_{sc}$, the upward trend in the SHAP value for $f_c'$ also slows down as $f_c'$ increases, indicating a gradual reduction in the impact of increasing $f_c'$ on the model output within larger ranges of $α_{sc}$. In situations with higher steel ratios, the contribution of increasing concrete strength to the model's predictive results gradually diminishes.

Observing the surface plot in Fig. 7(c), for SHAP value of $α_{sc}$, a trend of initially increasing and then decreasing with the increase of $α_{sc}$ is evident within the range of $f_c'$ from 0 to 200 MPa. This implies that within smaller ranges of $α_{sc}$, increasing the steel ratio significantly enhances the load-bearing capacity of the structure. However, when αsc exceeds a certain threshold, increasing the steel ratio may negatively impact the load-bearing capacity. This indicates the existence of an optimum $α_{sc}$ for different concrete strengths, maximizing its contribution to load-bearing capacity. As $f_c'$ increases from 0 to 200, these threshold values decrease from 0.26 to 0.10. This implies that with the increase of concrete strength, the most favorable steel ratio for enhancing the model output decreases.

Based on the observations and analyses of Fig. 7(a), Fig. 7(b) and Fig. 7(c), the following conclusions and design recommendations are derived: (1) When designing CFST columns, the nonlinear relationship between concrete compressive strength and steel ratio needs to be considered comprehensively. Choosing an appropriate combination of $f_c'$ and $α_{sc}$ can significantly enhance the load-bearing capacity of the structure, as indicated in Table 9. (2) Recognition of saturation effects: As concrete strength increases, the positive impact on load-bearing capacity may gradually diminish, leading to a saturation state. Similarly, an optimal saturation state is reached with increasing steel ratio, exhibiting the highest positive impact on load-bearing capacity. (3) In high-strength concrete, a lower steel ratio is required to achieve optimal load-bearing performance. These design recommendations will contribute to optimizing the performance of concrete structures and achieving the best load-bearing effects under different conditions.

Table 9 Design guidance for combination of $f_c'$ and $α_{sc}$

| $α_{sc}$ | $f_c'$ (MPa) | | | | | |
|---|---|---|---|---|---|---|
| | 0 | 40 | 80 | 120 | 160 | 200 |
| | 0.25 | 0.22 | 0.19 | 0.16 | 0.13 | 0.1 |

# 4. Conclusion

This project laid the groundwork for domain-adapted NNs. The key findings can be summarized as follows:

1. The introduction of domain-adapted knowledge has enhanced NNs, enabling proficient estimation of the load-bearing capacity of circular CFSTs while considering a wide spectrum of geometric and mechanical properties. These properties fall within the following ranges: cross-section diameter ranging from 44.95mm to 1020mm, steel tubular thickness from 0.52mm to 30mm, cross-section height from 114.3mm to 5560mm, concrete cylinder compressive strength from 6.41 MPa to 200.00 MPa, and steel yield strength from 178.28 MPa to 1153 MPa.

2. The models showcased high accuracy in predicting the load-bearing capacity of circular CFST columns, evidenced by 91.810% of predictions having an error margin below 20% and a MAPE value of 7.844%.

3. A notable correlation between the model predictions and experimental data was observed, surpassing the performance of existing codes and models. The models' scope, encompassing a more extensive range of geometric and mechanical properties, substantially enhances their practical applicability in CFST design.

4. Sensitivity analysis identified the diameter, circumference, and the cross-sectional areas of both the steel tube and the concrete core as the primary influencing factors in the model's performance.

5. SHAP analysis revealed that there are saturation effects for the materials. As the strength of core concrete and steel tube increases, the positive impact on load-bearing capacity may gradually diminish, leading to a saturation state. Additionally, choosing an appropriate combination of fc' and $α_{sc}$ can significantly enhance the structure's load-bearing capacity.

6. Robustness experiments indicated that the domain-adapted NNs exhibit greater stability and robustness compared to baseline models, maintaining acceptable error levels even under conditions where approximately 50% of the data is noisy and the error magnitude is up to 30%.

CRediT authorship contribution statement

Dian Wang: Data curation, Formal analysis, Writing - original draft. Gen Kondo: Conceptualization, Methodology, Formal analysis, Writing - original draft, review & editing. Zhigang Ren: Conceptualization, Methodology, Supervision, Writing - review & editing. Peipeng Li: Conceptualization, Methodology, Writing - review & editing.

Declaration of Competing Interest

The authors declare that they have no known competing financial interests or personal relationships that could have appeared to influence the work reported in this paper.

Acknowledgements

This research was financially supported by the National Natural Science Foundation of China (Grant No. 52108170) and the China Scholarship Council.

Code Availability: The model and datasets developed and analyzed in this study are openly available in the GitHub repository(https://github.com/kondogen/DKNN-model-for-of-circular-CFST-axial-capacity-prediction). Further materials related to this research may be obtained upon request.

Appendix A. Supplementary data: Containing supplementary data, is provided in a separate document available as an attachment.